\documentclass[twocolumn,prb,superscriptaddress,noshowpacs,epsf]{revtex4}

\usepackage{graphicx}
\usepackage{graphicx}
\usepackage{epsf}
\usepackage{amsmath}
\usepackage[dvips]{color}

\begin{document}

\newcommand{\chk}[1]{{\textcolor{red}{#1}}}

\title{Quantum emulation of a spin system with topologically protected ground states using superconducting quantum circuits}
\date{\today}
\author{J. Q. You}
\affiliation{Department of Physics and Surface Physics Laboratory (National Key Laboratory), Fudan University, Shanghai 200433, China}
\affiliation{Advanced Study Institute, The Institute of Physical and Chemical Research (RIKEN), Wako-shi 351-0198, Japan}
\author{Xiao-Feng Shi}
\affiliation{Department of Physics and Surface Physics Laboratory (National Key Laboratory), Fudan University, Shanghai 200433, China}
\affiliation{Advanced Study Institute, The Institute of Physical and Chemical Research (RIKEN), Wako-shi 351-0198, Japan}
%
\author{Xuedong Hu}
\affiliation{Advanced Study Institute, The Institute of Physical and Chemical Research (RIKEN), Wako-shi 351-0198, Japan}
\affiliation{Department of Physics, University at Buffalo, SUNY, Buffalo, NY 14260-1500, USA}
\author{Franco Nori}
\affiliation{Advanced Study Institute, The Institute of Physical and Chemical Research (RIKEN), Wako-shi 351-0198, Japan}
\affiliation{Center for Theoretical Physics, Physics Department,
University of Michigan, Ann Arbor, MI
48109-1040, USA}

\begin{abstract}
Using superconducting quantum circuit elements, we propose an approach to experimentally construct a Kitaev lattice, which is an anisotropic
spin model on a honeycomb lattice with three types of nearest-neighbor interactions and having topologically protected ground states.  We study
two particular parameter regimes to demonstrate both vortex and bond-state excitations.  Our proposal outlines an experimentally realizable
artificial many-body system that exhibits exotic topological properties.
\end{abstract}
\maketitle

\section{Introduction}
%
%
Interesting phenomena, such as the Aharonov-Bohm effect and Berry
phases, can occur in physical systems with nontrivial topology in
real or parameter space.  Topological quantum systems are now
attracting considerable interest because of their fundamental
importance in diverse areas ranging from quantum field theory to
semiconductor physics,\cite{RMP} with the most recent example being
the exploration of topological insulators.\cite{Kane_PRL05,Bernevig}
These topological physical systems may also have potential
applications because they are robust against local perturbations.
Specifically, a topologically protected quantum state degeneracy
cannot be lifted by any local interactions.\cite{RMP,Wen} It is
therefore natural to consider using topological phases for
applications requiring a high degree of quantum coherence.\cite{RMP}
For example, it has recently been pointed out that non-Abelian
anyons\cite{Leinaas,Wilczek,Wilczek_book} in a fractional quantum
Hall system can lead to topological quantum computing.\cite{Sarma05}
Anyons are neither bosons nor fermions, but obey anyonic braiding
statistics.\cite{Leinaas,Wilczek,Wilczek_book} Unfortunately, they
have not yet been convincingly observed experimentally in any
physical system.

%
%
Instead of only looking for naturally existing topological phases, one could also design artificial lattice structures that possess desired
topological phases.  One example is the Kitaev honeycomb model,\cite{Kitaev} which requires that the spin (natural or artificial) at each node
of a honeycomb lattice interacts with its three nearest neighbors through three different interactions: $\sigma_{x}\sigma_{x}$, $\sigma_{y}
\sigma_{y}$, and $\sigma_{z}\sigma_{z}$.  Depending on the bond parameters, this anisotropic spin model supports both Abelian and non-Abelian
anyons.~\cite{Kitaev}  Its realization could potentially lead to experimental demonstration of anyons and implementation of topological quantum
computing. However, the requirement for anisotropic interactions is tremendously demanding and generally cannot be satisfied by natural spin
lattices.

%
%
Various artificial lattices may possess interesting topological phases.  For instance, it has been proposed that a triangular Josephson junction
array may have a two-fold degenerate ground state that is topologically protected.\cite{Ioffe_Nat02,Albuquerque_PRB08} A recent proposal
suggests the use of capacitively coupled Josephson junction arrays to simulate a two-component fermion model that has topological
excitations.\cite{Xue_PRA09} There is also a suggestion that a Josephson junction array with properly designed interactions and topology can be
local-noise resistant.\cite{Gladchenko} With respect to the physical realization of the Kitaev model, there are proposals using neutral atoms in
optical lattices.\cite{Duan,Zoller,Sarma07}  One similarity among all of these proposals, whether based on Josephson junction arrays or on
optical lattices, is that they all require extremely low temperatures and precise single-atom manipulations.  The reason is that topologically
interesting properties are not generally contained in the symmetry of the system Hamiltonian.  Instead they are only emergent properties at very
low temperatures.

%
%
Here we propose a quantum emulation of the Kitaev lattice using
superconducting quantum circuits (Ref.~\onlinecite{YSN_preprint}
gives a brief summary of this work). As for the topic of quantum
analog simulations, see Ref.~\onlinecite{Buluta} for a brief review.
In our superconducting network, a Josephson charge qubit is placed
at each node of a honeycomb lattice.  These charge qubits behave
like artificial spins and are tunable via external
fields.\cite{YN05,Mak,Wendin} Each charge qubit interacts with its
three nearest neighbors through three different types of circuit
elements. One advantage of our proposal is that some circuit
elements involved and their functionalities at low energies have
already been demonstrated experimentally---for example, the
$\sigma_{z}\sigma_{z}$ and $\sigma_{x} \sigma_{x}$ couplings between
charge qubits have been studied in experiments.\cite{NEC,Yamamoto}
Here we show theoretically that they can indeed provide the needed
anisotropic interactions when included in a honeycomb lattice. We
then identify the ground states of this network in two different
parameter regimes and show that it can have both vortex and
bond-state excitations. We also describe how they can be generated
using spin-pair operations.

\begin{figure}
\includegraphics[width=3in,
%
bbllx=100,bblly=213,bburx=467,bbury=725] {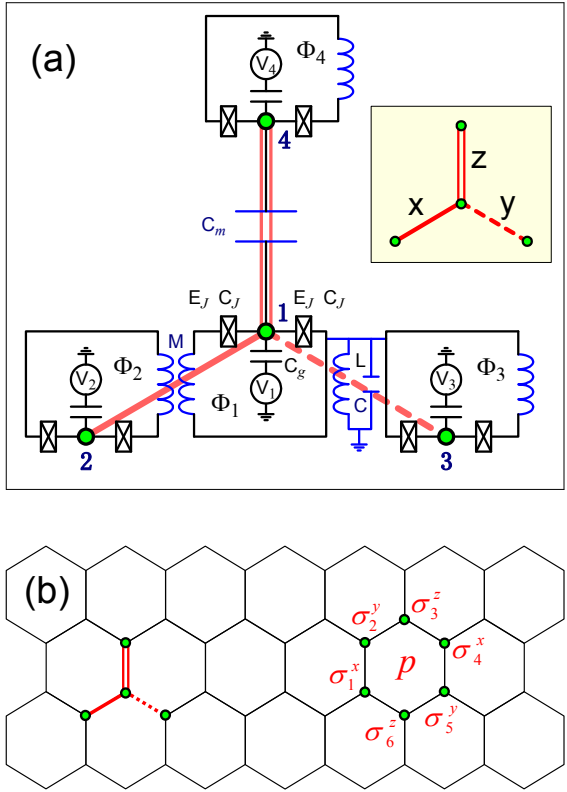} \caption{(Color
online)(a)~Schematic diagram of the basic building block of a Kitaev
lattice, consisting of four superconducting charge qubits (labelled
1 to 4): (i)~Qubits 1 and 2 are inductively coupled via a mutual
inductance $M$; (ii)~qubits 1 and 3 are coupled via an $LC$
oscillator; and (iii)~qubits 1 and 4 are capacitively coupled via a
mutual capacitance $C_m$. Inset: The three types of inter-qubit
couplings are denoted as $x$-, and $y$- or $z$-type bonds. Here each
charge qubit consists of a Cooper-pair box (green dot) that is
linked to a superconducting ring via two identical Josephson
junctions (each with coupling energy $E_J$ and capacitance $C_J$),
to form a SQUID loop. Also, each qubit is controlled by both a
voltage $V_i$ (applied to the qubit via the gate capacitance $C_g$)
and a magnetic flux $\Phi_i$ (piercing the SQUID loop). (b) A
partial Kitaev lattice (honeycomb lattice) constructed by repeating
the building block in (a), where a charge qubit is placed at each
site.  A plaquette is defined as a hexagon in the lattice.  The
plaquette operator is defined as $W_p = \sigma_1^x \sigma_2^y
\sigma_3^z \sigma_4^x \sigma_5^y \sigma_6^z$ and is shown for a
given plaquette $p$.} \label{fig1}
\end{figure}

\section{Kitaev lattice based on superconducting quantum circuits}

At low energies, superconducting (SC) qubits can behave as
artificial spins.  Among the varieties of SC qubits (charge, flux,
phase,\cite{YN05,Mak,Wendin} and other
hybrids~\cite{YTN06,YHAN07,transmon}), only charge qubits are known
to interact with each other in all the individual forms of
$\sigma_{x}\sigma_{x}$, $\sigma_{y}\sigma_{y}$, and
$\sigma_{z}\sigma_{z}$ (via a mutual inductance, an $LC$ oscillator,
and a capacitance, respectively).\cite{YTN,Schon,NEC,Bruder}
Therefore, to emulate a Kitaev honeycomb lattice, we propose to
build a two-dimensional SC circuit network based on SC charge
qubits. More specifically, on a honeycomb lattice a charge qubit is
placed on each node [Fig.~\ref{fig1}(b)], and one of the three
circuit elements is inserted along each bond of the lattice (denoted
as the $x$-, $y$-, or $z$-type bond). A building block of this
lattice is shown in Fig.~\ref{fig1}(a), which consists of four
charge qubits that are connected via an $x$-, a $y$-, and a $z$-type
bond. Each charge qubit is a Cooper-pair box connected to a
superconducting ring by two identical Josephson junctions to give it
tunability: Each qubit is controlled by both the magnetic flux
$\Phi_i$ piercing the SQUID loop and the voltage $V_i$ applied via
the gate capacitance $C_g$.

Naively, a circuit element should maintain its basic characteristics
when inserted in a larger network, at least in the linear regime.
However, as it has been shown in previous studies of hybrid
qubits,\cite{YTN06,YHAN07,transmon} a superconducting qubit based on
one particular variable (for example, charge) can acquire characters
of another (for example, flux) when additional circuit elements are
added to it.  Therefore, here we first clarify whether the different
circuit elements in our honeycomb network maintain their basic
individual characteristics (particularly the forms and strengths of
the interactions) at low energies when lumped together.

We first write down the Lagrangian of the quantum circuits, choosing
the average phase drop $\varphi_i$ across the two Josephson
junctions of each charge qubit as the canonical coordinates.  After
identifying the corresponding canonical momenta, we then derive
(this derivation is shown in the appendix) the total Hamiltonian of
the quantum circuits as
\begin{eqnarray}
H\!&\!=\!&\!\sum_i H_i + \sum_{x-{\rm link}}K_x(j,k) + \sum_{y-{\rm
link}}K_y(j,k) \nonumber \\
&&\!+\sum_{z-{\rm link}}K_z(j,k).
\label{hamiltonian}
\end{eqnarray}
Here the free Hamiltonian of the $i$th charge qubit is
\begin{equation}
H_i=E_c(n_i-n_{gi})^2 - E_{Ji}(\Phi_i)\cos\varphi_i,
\end{equation}
where $E_c = 2e^2 C_\Sigma/\Lambda$ is the charging energy of the
Cooper pair box, with the total capacitance $C_{\Sigma} =
2C_J+C_g+C_m$, and $\Lambda = C_{\Sigma}^2 -C_m^2$; $n_i =
-i\partial/\partial\varphi_i$ the number operator of the Cooper
pairs in the $i$th box (which is conjugate to $\varphi_i$); $n_{gi}
= C_g V_i/2e$ the reduced offset charge induced by the gate voltage
$V_i$; and $E_{Ji}(\Phi_i) = 2E_J \cos(\pi\Phi_i/\Phi_0)$ the
effective Josephson coupling energy of the $i$th charge qubit, with
$\Phi_0=h/2e$ the flux quantum.

The three nearest-neighbor couplings, shown as the $x$, $y$, and $z$ bonds in Fig.~1(a), are given by
\begin{eqnarray}
K_x(1,2)\!&\!=\!&\!M I_1 I_2, \nonumber\\
K_y(1,3)\!&\!=\!&\! -4 \xi E_{J1}(\Phi_1) E_{J3}(\Phi_3) \sin\varphi_1
\sin\varphi_3,\\
K_z(1,4)\!&\!=\!&\!E_m (n_1 - n_{g1}) (n_4 - n_{g4}),\nonumber
\end{eqnarray}
where
\begin{eqnarray}
&&\xi=L \left[\frac{\pi C_{\Sigma} (C_g + C_m)}{\Lambda\Phi_0}\right]^2, \nonumber\\
&&E_m =\frac{4e^2 C_m}{\Lambda}, \nonumber \\
&&I_i = -I_c \sin\left(\frac{\pi\Phi_i}{\Phi_0}\right) \cos\varphi_i \,.
\end{eqnarray}
Here $I_c = 2\pi E_J/\Phi_0$ is the critical current through the Josephson junctions of the charge qubits (we assume identical junctions for
simplicity), while $I_i$ is the circulating supercurrent in the SQUID loop of the $i$th charge qubit. Note that the coupling strength between
nodes 1 and 3 (along a $y$-link), $\xi \propto (C_g + C_m)^2$, is affected by the mutual capacitance $C_m$ that connects qubit 1 (3) with its
nearest-neighbor along a $z$-link. Compared to the case of two qubits coupled by an $LC$ oscillator,\cite{Schon} where $\xi \propto C_g^2$, the
capacitive inter-node coupling along the $z$-link in the present circuit greatly increases the inter-node coupling along the $y$-link because
usually $C_m \gg C_g$. This is an important and positive consequence when multiple circuit elements are introduced to create different
inter-node interactions.

At low temperatures, only the lowest-energy states of a superconducting circuit element are involved in the system dynamics, which is quantum
mechanical.  For the particular case of a charge qubit, where $E_c \gg E_J$, the lowest-energy eigenstates are mixtures of having zero and one
Cooper pair in the box, when the gate voltage $V_i$ is near the optimal point $e/C_g$ (i.e., $n_{gi}\sim \frac{1}{2}$).  Defining $|0\rangle_i$
and $|1\rangle_i$ as the two charge states having zero and one extra Cooper pair in the box, we now have a two-level system as a quantum bit, or
qubit.  In the spin-$\frac{1}{2}$ representation based on these charge states $|0\rangle_i\equiv|\!\!\uparrow\rangle_i$ and $|1\rangle_i \equiv
|\!\!\downarrow\rangle_i$ ($i$ is the index of the nodes), the system variables can be expressed as
\begin{eqnarray}
&&n_i = \frac{1}{2}(1 -\sigma_i^z),\nonumber\\
&&\cos\varphi_i=\frac{1}{2}\sigma_i^x,\\
&&\sin\varphi_i =-\frac{1}{2}\sigma_i^y.\nonumber
\end{eqnarray}
Here we consider the simple case with $n_{gi} = n_g$ (i.e., all gate voltages on the different nodes are identical: $V_i=V_g$) and $\Phi_i =
\Phi_e$ for all qubits.  The low-energy Hamiltonian of the system is then reduced to
\begin{eqnarray}
H\!&\!=\!&\!J_x\sum_{x-{\rm link}}\sigma_j^x\sigma_k^x + J_y\sum_{y-{\rm link}}\sigma_j^y\sigma_k^y +J_z\sum_{z-{\rm link}}
\sigma_j^z\sigma_k^z \nonumber \\
&&\!+\sum_i(h_z\sigma_i^z+h_x\sigma_i^x),
\label{model}
\end{eqnarray}
where
\begin{eqnarray}
J_x\!&\!=\!&\!\frac{1}{4}MI_c^2\sin^2\left(\frac{\pi
\Phi_e}{\Phi_0}\right)\geq 0,\nonumber\\
J_y\!&\!=\!&\!-\xi[E_{J}(\Phi_e)]^2\leq 0, \nonumber \\
J_z\!&\!=\!&\!\frac{1}{4}E_m>0,\\
h_z\!&\!=\!&\!\left(E_c+\frac{1}{2}E_m\right)\left(n_{g}-\frac{1}{2}\right), \nonumber \\
h_x\!&\!=\!&\!-\frac{1}{2}E_{J}(\Phi_e),\nonumber
\end{eqnarray}
with $E_{J}(\Phi_e) = 2E_J\cos(\pi\Phi_e/\Phi_0)$.  The reduced
Hamiltonian (\ref{model}) is the Kitaev model with an effective
magnetic field with $z$- and $x$-components.  Here $h_x$ and $h_z$
play the role of a ``magnetic'' field.  Since $J_y\propto h_x^2$, to
maintain finite inter-qubit couplings, $h_x$ cannot vanish.
Therefore our Hamiltonian represents a Kitaev model in an
always-finite magnetic field, although the field direction can be
adjusted.  This Hamiltonian has an extremely complex quantum phase
diagram because of all the (experimentally) adjustable parameters.
Here we are particularly interested in whether it has
topologically-interesting phases and when such topological
properties might emerge.

\section{Vortex and bond-state excitations}

Below we focus on two particular parameter regimes of the
finite-field Kitaev model of (\ref{model}), under the general
condition that the $z$-bond interaction dominates over the other
interactions. In particular, when $J_z \gg J_x, |J_y| \gg |h_z|,
|h_x|$, we identify a vortex state excitation. This case is
described in Section A below.  When $h_z = 0$ but $h_x$ is of the
same order as $J_x$ and $J_y$, we identify a new excitation that we
call the bond state.  We describe this case in Sec.~III.B.  The
vortex state is a known topological excitation in the zero-field
Kitaev model, while the bond state is specific to the finite-field
Kitaev model.

\subsection{Kitaev lattice with dominant $z$-bonds in a weak ``magnetic'' field}

We first consider the case when
\begin{equation}
J_z \gg J_x, |J_y| \gg |h_z|, |h_x|,
\end{equation}
and treat $V=\sum_i(h_z\sigma_i^z+h_x\sigma_i^x)$ as the perturbation.  Using perturbation theory in the Green function formalism,~\cite{Kitaev}
one can construct an effective Hamiltonian for the lattice:
\begin{equation}
H' = -\frac{2h_z^2}{\Delta\varepsilon_z} \sum_{z-{\rm link}} \sigma_j^z \sigma_k^z - \frac{2h_x^2}{\Delta\varepsilon_x} \sum_{x-{\rm link}}
\sigma_j^x \sigma_k^x,
\end{equation}
where $\Delta\varepsilon_{z(x)}$ is the excitation energy of the
state $\sigma_i^{z(x)}|g_0\rangle$, i.e., the energy difference
between states $\sigma_i^{z(x)}|g_0\rangle$ and $|g_0\rangle$. Here
$|g_0\rangle$ is the ground state of the unperturbed Hamiltonian,
i.e., Hamiltonian (\ref{model}) with the perturbation term $V$
excluded.  Note that the effective Hamiltonian $H'$ only contains
contributions from the second-order terms because both the first-
and third-order terms vanish.  Including the zeroth-order term
(unperturbed Hamiltonian), the total Hamiltonian of the system can
be written as
\begin{equation}
H = J'_x \sum_{x-{\rm link}} \sigma_j^x \sigma_k^x + J_y
\sum_{y-{\rm link}} \sigma_j^y \sigma_k^y + J'_z \sum_{z-{\rm link}}
\sigma_j^z \sigma_k^z \,,
\end{equation}
where the effective $z$- and $x$-couplings are
\begin{eqnarray}
J'_z \!&\!=\!&\! J_z - \frac{2h_z^2}{\Delta \varepsilon_z}, \nonumber\\
J'_x \!&\!=\!&\! J_x - \frac{2h_x^2}{\Delta \varepsilon_x}\,.
\end{eqnarray}

\begin{figure}
\includegraphics[width=2.6in,
bbllx=145,bblly=445,bburx=418,bbury=745] {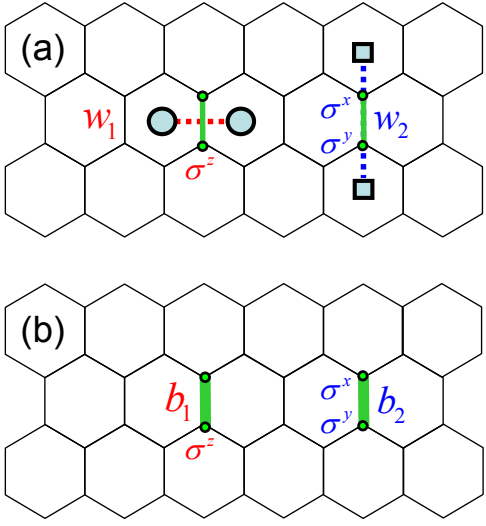} \caption{(Color online) (a)~Two types of vortex excitations $w_1$ and $w_2$. A pair of
vortices are generated along the horizontal direction for $w_1$ (vertical direction for $w_2$) by the spin-pair operator
$\tilde{\sigma}^z\equiv\sigma^zI$ ($\tilde{\sigma}^y\equiv\sigma^y\sigma^x$) acting on a $z$-link. (b)~Two bond-state excitations $b_1$ and
$b_2$, which are also generated by the spin-pair operators $\tilde{\sigma}^z$ and $\tilde{\sigma}^y$ on a $z$-link. } \label{fig2}
\end{figure}

Below we focus on the Abelian excitations.  When $J'_z \gg J'_x,
|J_y|$, the dominant part of the Hamiltonian $H$ is that along the
vertical links,
\begin{equation}
H_0 = J'_z\sum_{z-{\rm link}} \sigma_j^z\sigma_k^z.
\end{equation}
Under $H_0$, the two spins along each $z$-link tend to be aligned opposite to each other ($|\!\!\uparrow \downarrow \rangle$ or $|\!\!\downarrow
\uparrow \rangle$) in order to lower their energies.  Indeed, the highly degenerate ground state $|g\rangle$ of $H_0$ is an arbitrary vector in
the Hilbert subspace spanned by $\bigotimes_{i=1}^N |\sigma \bar{\sigma} \rangle_i$, where $N$ denotes the total number of $z$-links and $\sigma
= \;\uparrow, \downarrow$.  Within the ground-state subspace of $H_0$ and up to fourth order,\cite{Kitaev} the effective Hamiltonian of the
Kitaev lattice takes the form
\begin{equation}
H_{\rm eff}=-J_{\rm eff}\sum_p W_p, \label{vortex}
\end{equation}
where
\begin{eqnarray}
J_{\rm eff} & = & \frac{{J'}_x^2 J_y^2}{16{J'}_z^3}\approx\frac{J_x^2 J_y^2}{16J_z^3},
\nonumber \\
W_p & = & \sigma_1^x \sigma_2^y \sigma_3^z \sigma_4^x \sigma_5^y
\sigma_6^z \,.
\end{eqnarray}
Here $W_p$ is the plaquette operator for a given plaquette $p$ [see Fig.~\ref{fig1}(b)].  The operator $W_p$ for any plaquette $p$ commutes with
the unperturbed Hamiltonian $H_0$:
\begin{equation}
[H_0, W_p] = 0;
\end{equation}
so that $[H_0, H_{\rm eff}] = 0$ as well, and the ground states $|g\rangle_w$ of $H_{\rm eff}$ form a {\it subset} of the degenerate ground
states ${|g\rangle}$ of $H_0$.  It is straightforward to show that $W_p^2 |g\rangle = |g\rangle$, or $W_p |g\rangle = \pm |g\rangle$.  Since
$J_{\rm eff} > 0$, to minimize the energy of $|g\rangle_w$, we need
\begin{equation}
W_p|g\rangle_w=|g\rangle_w.
\end{equation}
In other words, the eigenvalues of the $W_p$ operators in the ground state $|g\rangle_w$ are $w_p = 1$ for all plaquettes $p$.

When some plaquettes undergo transformations that lead to $w_p=1\rightarrow-1$, the system gets into an excited state.  The lowest-energy
excitation corresponds to the generation of a pair of vortices when $w_p=1\rightarrow-1$ for two neighboring plaquettes.  In this excitation
process, each of the two neighboring plaquettes acquires a phase $\pi$, which is equivalent to the addition of a flux quantum $\Phi_0$ through
each plaquette. As shown in Fig.~2(a), such an excitation can be generated by applying either of the following two spin-pair operators on the
ground state $|g\rangle_w$:
\begin{equation}
|\widetilde{Z}_i\rangle = \tilde{\sigma}_i^z|g\rangle_w,~~~
|\widetilde{Y}_i\rangle = \tilde{\sigma}_i^y|g\rangle_w,
\end{equation}
with
\begin{equation}
\tilde{\sigma}_i^z\equiv\sigma_{i}^zI_{i},~~~
\tilde{\sigma}_i^y\equiv\sigma_{i}^y\sigma_{i}^x.
\end{equation}
Here the two operators $\sigma^z_i$ ($\sigma^y_i$) and $I_i$ ($\sigma^x_i$) act on the ground state $|g\rangle_w$ at the bottom and top sites of
the $i$th $z$-link, respectively.  This pair of vortices, generated by either $\tilde{\sigma}_i^z$ or $\tilde{\sigma}_i^y$, are topological
states with an excitation energy of
\begin{equation}
\Delta\varepsilon=4J_{\rm eff}
\end{equation}
above the ground state.  As shown in Refs.~\onlinecite{Kitaev} and \onlinecite{Sarma07}, these excitations exhibit the braiding statistics of
Abelian anyons.  The ratio between this excitation gap for the anyons and $J_z$ is
\begin{equation}
\frac{\Delta\varepsilon}{J_z} \sim \left(\frac{J_x J_y}{J_z^2}\right)^2 \ll 1.
\end{equation}
For example, if $J_z \sim 10$GHz and both $J_x$ and $J_y$ are one tenth of $J_z$, this gap would be about 1 MHz, corresponding to a temperature
of 0.1 mK.  This small gap requires an extremely low experimental temperature for suppressing the thermal activation of the ground state to the
vortex states.  Note that a different perturbative approach~\cite{Vidal} shows that in the parameter region $J'_z \geq J'_x, |J_y|$, the
spin-pair operators $\tilde{\sigma}^z_i$ and $\tilde{\sigma}^y_i$ generally create both vortex and fermionic excitations.  However, in the limit
of $J'_z \gg J'_x, |J_y|$, the dominant excitations are vortex states,~\cite{Vidal} which is consistent with the conclusion drawn above.

\subsection{Kitaev lattice with dominant $z$-bonds in a uniform ``magnetic'' field along the $x$-direction}

If we stay in the regime where the $z$-bond couplings are dominant
($J_z \gg J_x, |J_y|$), but place each charge qubit at the optimal
point where $n_g=\frac{1}{2}$, so that $h_z=0$, a different quantum
phase arises when $|h_x|$ is comparable to $J_x, |J_y|$.  In other
words, we now consider the regime
\begin{equation}
J_z \gg J_x, |J_y|, |h_x|, \ \ {\rm and} \ h_z = 0 \,.
\end{equation}
Here the zeroth-order Hamiltonian is again the coupling along the $z$-bonds: $H_0 = J_z \sum_{z-{\rm link}} \sigma_j^z \sigma_k^z$ (notice that
here the coupling strength is $J_z$, not $J'_z$), with the same highly degenerate ground state $|g\rangle$ as discussed in the previous
subsection. To clarify the low-energy excitation spectrum in this regime, we again use perturbation theory in the Green's function formalism to
remove the linear terms and derive an effective Hamiltonian in the ground state sub-Hilbert space of $H_0$.  Up to second-order, the effective
Hamiltonian takes the form
\begin{equation}
H_{\rm eff}^{(z)} = -K_{\rm eff} \sum_{z-{\rm link}} \sigma_j^x \sigma_k^x,
\label{bond}
\end{equation}
where
\begin{equation}
K_{\rm eff} = \frac{h_x^2}{J_z}.
\end{equation}
The spin pair operator $P_z=\sigma_j^x\sigma_k^x$ at a $z$-bond
(again the two Pauli operators act on the bottom and top nodes of
the particular $z$-bond) commutes with the zeroth-order Hamiltonian
$[P_z, H_0] = 0$ (although it anti-commutes with the four plaquette
operators $W_p$ connected to this $z$-bond).  Similar to $W_p$, the
pair operator $P_z$ also has two eigenvalues $p_z = \pm 1$. Thus the
ground state $|g\rangle_b$ of $H_{\rm eff}^{(z)}$ should satisfy
$p_z=1$ for all the $z$-bonds in the system.  In other words,
\begin{equation}
P_z|g\rangle_b=|g\rangle_b,
\end{equation}
for all $z$-bonds.  Since no two $z$-bonds share a node in the honeycomb lattice, and the lattice is completely covered by all the $z$-bonds, we
can solve the eigenstates of $P_z$ of each $z$-bond and obtain the ground state of $H_{\rm eff}^{(z)}$ as
\begin{equation}
|g\rangle_b = \frac{1}{2^{N/2}} \bigotimes_{i=1}^N(|\!\!\uparrow\downarrow\rangle_i+ |\!\!\downarrow\uparrow\rangle_i).
\end{equation}
This is a nondegenerate ground state, which forms a simple subset of the highly degenerate ground states ${|g\rangle}$ of $H_0$.  It is
maximally entangled within each $z$-bond, but not entangled at all between different $z$-bonds.  In other words, the two-spin correlation
function decays to identically zero beyond a $z$-bond. The lattice is now an ensemble of maximally entangled ``spin'' pairs that are completely
independent from each other.  This ground state is reminiscent of (and simpler than) the dimerized valence bond solid state discussed in the
context of spin Hamiltonians.\cite{Affleck_87,Majumdar_69}  There valence bond states refer to a singlet $|\uparrow\downarrow
-\downarrow\uparrow\rangle$ for the electron spins, which is dictated by the Coulomb interaction and Pauli principle between electrons.

When the pair operators $\tilde{\sigma}_i^z$ and $\tilde{\sigma}_i^y$ are separately applied to the ground state at the $i$th $z$-bond [see
Fig.~2(b)], the excited states
\begin{equation}
|\widetilde{Z}_i\rangle=\tilde{\sigma}_i^z|g\rangle_b,~~~
|\widetilde{Y}_i\rangle=\tilde{\sigma}_i^y|g\rangle_b
\end{equation}
are called a bond state---while the pair operators are different,
the states they generate are only different by an overall phase
because $|g\rangle_b$ is a factored state for all $z$-bonds.  A bond
state at the $i$th $z$-bond corresponds to the change of
$p_z=1\rightarrow -1$ at that particular bond.  It is $2 K_{\rm
eff}$ above the ground state in energy.  Notice that a bond state is
an excitation that is completely localized to a particular $z$-bond.
Furthermore, bond states are generated by the same pair operators
that generate the vortex excitations, although the ground states of
the system are different in these two cases.  In contrast to
$|g\rangle_w$, the ground state $|g\rangle_b$ is {\it
nondegenerate}, and the bond state excitations are very different
from the vortex states.  This transition from vortex excitations to
bond states occurs when we vary the parameters of the system (i.e.,
tuning $n_g$ to $1/2$ and reduce $\Phi_e$ from close to $\Phi_0/2$
so that $h_x$ increases to the same magnitude as $J_x$ and/or
$J_y$), during which the topological property of the system changes.

\begin{figure}
\includegraphics[width=3in,  
bbllx=136,bblly=217,bburx=494,bbury=742]{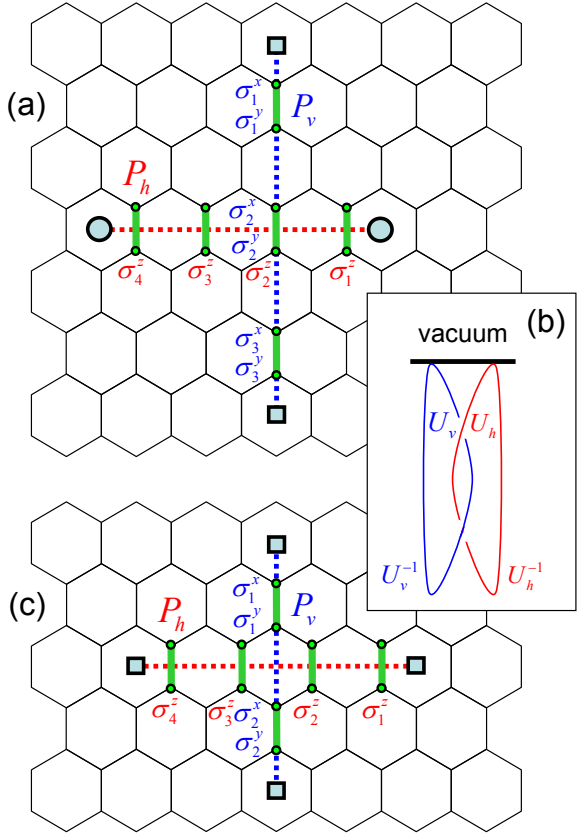} \caption{(Color
online) Schematic diagram of the procedures for braiding
excitations. (a)~The operations $U_h$ and $U_v$ for creating
excitations, which are achieved by successively applying spin-pair
operators at $z$-bonds along the horizontal ($P_h$) and vertical
($P_v$) paths. Here the paths $P_v$ and $P_h$ intersect at a
$z$-bond. (b)~A combined operation $U_h^{-1}U_v^{-1}U_hU_v$ for
both, braiding the excitations created in (a), and fusing them to
the vacuum. (c)~The operations $U_h$ and $U_v$ for creating
excitations, which are also achieved by successively applying
spin-pair operators at $z$-bonds along $P_h$ and $P_v$, but the
paths $P_v$ and $P_h$ do not intersect at a $z$-bond.} \label{fig3}
\end{figure}

\section{The braiding of excitations}

A vortex looping around another vortex can produce either a sign
change or {\it no} sign change to the wave function.  The first case
is denoted as an $e$-type vortex looping around an $m$-type vortex,
and the second case corresponds to an $e$-type vortex looping around
an $e$-type vortex (see, e.g., Ref.~\onlinecite{Sarma07} for a more
detailed discussion).  This indicates anyonic statistics between the
$e$ and $m$ vortex states. Therefore, braiding, which refers to
moving one quasi-particle around another, is an important tool to
determine the statistics of the quasi-particles (in the present case
the vortices). Here we show an alternative procedure for braiding an
excitation with another, which can be applied to both vortex and
bond states.

Let us consider two particular evolutions for the system.  The first evolution $U_v$ contains spin-pair operations $\tilde{\sigma}_i^y =
\sigma_{i}^y \sigma_{i}^x$ applied to the ground state $|\tilde{g}\rangle$ at three successive $z$-bonds along the vertical path $P_v$, as shown
in Fig.~\ref{fig3}(a).  Here $|\tilde{g}\rangle \equiv |g\rangle_w$ for the vortex case and $|\tilde{g}\rangle \equiv |g\rangle_b$ for the
bond-state case.  The second evolution $U_h$ contains spin-pair operations $\tilde{\sigma}_i^z=\sigma_{i}^zI_{i}$ applied at four successive
$z$-bonds along the horizontal path $P_h$ as shown in Fig.~\ref{fig3}(a).  After these two operations in series, the state of the system is $U_h
U_v |\tilde{g}\rangle$, where
\begin{equation}
U_h = \tilde{\sigma}_4^z\tilde{\sigma}_3^z\tilde{\sigma}_2^z\tilde{\sigma}_1^z, ~~~U_v=\tilde{\sigma}_3^y\tilde{\sigma}_2^y\tilde{\sigma}_1^y
\,.
\label{operation}
\end{equation}
Now we turn the evolutions backward by applying $U_v^{-1}$ and $U_h^{-1}$ to the system successively, so as to fuse~\cite{RMP,Kitaev} the
excitations to the vacuum (i.e., the ground state) [see Fig.~\ref{fig3}(b)].  The final state of the system is now
\begin{equation}
|\Psi_f\rangle=U_h^{-1}U_v^{-1}U_hU_v|\tilde{g}\rangle.
\label{braid}
\end{equation}
When the paths $P_v$ and $P_h$ intersect at a $z$-bond, such as in the example given in Fig.~\ref{fig3}(a), where $\tilde{\sigma}^y_2$ and
$\tilde{\sigma}^z_2$ anti-commute, $U_h$ and $U_v$ anti-commute as well: $U_hU_v=-U_vU_h$.  The final state thus becomes
\begin{equation}
|\Psi_f\rangle=-|\tilde{g}\rangle \,.
\end{equation}
In other words, a phase flip $e^{i\pi}$ resulted from the
evolutions.  For vortex excitations, this is equivalent to the case
of an $e$-type vortex looping around an $m$-type vortex, as shown in
Ref.~\onlinecite{Sarma07}. In contrast, when similar operations are
applied but the paths $P_v$ and $P_h$ do not intersect at a $z$-bond
[see Fig.~\ref{fig3}(c), for example], $U_h U_v = U_v U_h$, so that
\begin{equation}
|\Psi_f\rangle=|\tilde{g}\rangle,
\end{equation}
yielding no phase flip in the final state as compared to the initial state. For vortex excitations, this is equivalent to the case of an
$e$-type vortex looping around another $e$-type vortex.

The braiding of excitations, i.e. whether there is or there is no
phase flip, can be revealed by means of Ramsey-type
interference.~\cite{Sarma07,Pachos}  To achieve this, one can keep
the same $U_v$ as above, but use
\begin{equation}
U_h=(\tilde{\sigma}^z_4)^{\frac{1}{2}}(\tilde{\sigma}^z_3)^{\frac{1}{2}}
(\tilde{\sigma}^z_2)^{\frac{1}{2}}(\tilde{\sigma}^z_1)^{\frac{1}{2}},
\end{equation}
where
\begin{equation}
(\tilde{\sigma}^z_i)^{\frac{1}{2}}\equiv(\sigma^z_i)^{\frac{1}{2}}I_i \,,
\end{equation}
i.e., each $\sigma^z_i$ is replaced by half of the rotation. In the braiding case shown in Fig.~\ref{fig3}(a),
\begin{equation}
\tilde{\sigma}^y_2(\tilde{\sigma}^z_2)^{\frac{1}{2}}
=i(\tilde{\sigma}^z_2)^{-\frac{1}{2}}\tilde{\sigma}^y_2
\end{equation}
at the crossing point of paths $P_h$ and $P_v$. Thus,
\begin{eqnarray}
|\Psi_f\rangle\!&\!=\!&\!U_h^{-1}U_v^{-1}U_hU_v|\tilde{g}\rangle
=(\tilde{\sigma}^z_2)^{-\frac{1}{2}}[i(\tilde{\sigma}^z_2)^{-\frac{1}{2}}]|\tilde{g}\rangle \nonumber\\
\!&\!=\!&\!i(\tilde{\sigma}^z_2)^{-1}|\tilde{g}\rangle=i\tilde{\sigma}^z_2|\tilde{g}\rangle =i|\widetilde{Z}_2\rangle,
\end{eqnarray}
similar to the case with an $e$ vortex braiding with a superposition state of an $m$ vortex and the vacuum.~\cite{Sarma07}  However, in the case
without braiding [see Fig.~\ref{fig3}(c)],
\begin{equation}
|\Psi_f\rangle=U_h^{-1}U_v^{-1}U_hU_v|\tilde{g}\rangle=|\tilde{g}\rangle.
\end{equation}
Therefore, the braiding of excitations can be distinguished by verifying if an excited state $|\widetilde{Z}_2\rangle$ occurs at the crossing
point of paths $P_h$ and $P_v$.

While the vortex state described by Eq.~(\ref{vortex}) and the bond state described by Eq.~(\ref{bond}) are very different excitations, they
have similar braiding properties.  In the braiding procedure shown above, the system is initially in the vacuum (either $|g\rangle_w$ or
$|g\rangle_b$); after the braiding operations in Eq.~(\ref{braid}), the system is fused to the vacuum again, but with a sign change to the
ground-state wave function no matter which ground state the system starts with.  In order to distinguish the difference between the vortex and
bond-state excitations, one needs to focus on the intermediate steps of the braiding operations.  Take $U_v$ in Eq. (\ref{operation}) as an
example. When it is applied to $|g\rangle_w$, the spin-pair operator $\tilde{\sigma}_1^y$ in it first creates a pair of $e$ vortices, and then
the other spin-pair operations $\tilde{\sigma}_2^y$ and $\tilde{\sigma}_3^y$ successively move one vortex downward along the vertical path
$P_v$. The final state $U_v |g\rangle_w$ is also a pair of vortices, but the two vortices are separated by three $z$-bonds in the vertical
direction [see Fig.~\ref{fig3}(a)].  Importantly, this pair of vortices $U_v|g\rangle_w$ is {\it degenerate} with the pair of vortices
$\tilde{\sigma}_1^y|g\rangle_w$.  However, in sharp contrast to the vortex case, when $U_v$ in Eq. (\ref{operation}) is applied to
$|g\rangle_b$, each of the spin-pair operations $\tilde{\sigma}_i^y$, $i=1,2,3$ creates a bond state and the final state $U_v |g\rangle_b$ is
{\it nondegenerate} with the bond state $\tilde{\sigma}_1^y|g\rangle_b$.

\section{Implementation of quantum rotations}

As indicated in previous sections, single-qubit rotations are needed to create vortex and bond-state excitations, and to perform braiding
operations.  Below we show that these quantum rotations of individual qubits in the honeycomb lattice can be achieved via electrical and
magnetic controls.  The key is to reduce the coupling between a specific qubit and its neighboring qubits to such a degree that its single-qubit
dynamics dominates for a period of time.

%
%
To generate a $\sigma_z$ rotation at a particular charge qubit, we consider the following approach by controlling both the magnetic flux through
SQUID loops and the local electric field.  Specifically, when the magnetic flux in the SQUID loop of each charge qubit is set to
$\Phi_e=\Phi_0/2$, $h_x = 0$ and $J_y = 0$, so that the honeycomb lattice is now decoupled into a series of one-dimensional chains.  For a
charge qubit $E_c \gg E_J(\Phi_e)$, thus $E_c \gg J_x$.  We further assume that $E_c \gg E_m$, so that $E_c\gg J_z$ as well.  One can now shift
the gate voltage for a period of time $\tau$ at the $i$th lattice point far away from the usual working point $n_g \sim \frac{1}{2}$ of the
Kitaev lattice, so that the corresponding single-qubit energy $\delta h_z \sim E_c$ (instead of $\sim E_J$) is much larger than both $J_z$ and
$J_x$.  Such a parameter regime should be reasonably easy to achieve for a charge qubit.  This operation of shifting $n_g$ should yield a local
$z$-type rotation on the $i$th qubit:
\begin{equation}
R_i^z(\theta)=\exp[-i(\delta h_z\tau/\hbar)\sigma_i^z]\equiv\exp(-i\theta\sigma_i^z/2),
\end{equation}
where
\begin{equation}
\delta h_z=h_z(n_{gi})-h_z(n_g).
\end{equation}
When $\theta \equiv 2\delta h_z\tau/\hbar=\pi$ (where $n_{gi}>n_g$), $R_i^z(\pi)=-i\sigma_i^z$, so the $\sigma_i^z$ operation on the $i$th qubit
is given by
\begin{equation}
\sigma_i^z=e^{i\pi/2}R_i^z(\pi),
\end{equation}
while half of the rotation is
\begin{equation}
(\sigma_i^z)^{\frac{1}{2}}=e^{i\pi/4}R_i^z(\pi/2).
\end{equation}
The corresponding inverse rotations can be achieved by shifting the gate voltage to $n_{gi}<n_g$.

A $\sigma_x$ rotation of a particular charge qubit can be generated by a similar approach.  Specifically, when $n_g = \frac{1}{2}$ and
$\Phi_e=0$, one has $h_z=0$, $h_x=-E_J$, and $J_x=0$.  Again the honeycomb lattice is separated into a series of one-dimensional chains.  Here
we assume that $E_J \gg |J_y|,J_z$, achievable in this charge-qubit system, which allows us to perform a single-qubit rotation driven by $E_J$.
When $n_g = \frac{1}{2}$, for a time $\tau$ we switch off the flux in the SQUID loop of the $i$th qubit (the working point of this Kitaev
lattice is usually at $0<\Phi_e<\Phi_0/2$), producing a local $x$-type rotation on the $i$th qubit:
\begin{equation}
R_i^x(\theta)=\exp[i(\delta E_J\tau/\hbar)\sigma_i^x] \equiv \exp(i\theta\sigma_i^x/2),
\end{equation}
where
\begin{equation}
\delta E_J=E_J-\frac{1}{2}E_J(\Phi_e).
\end{equation}
The $\sigma_i^x$ rotation on the $i$th qubit is
\begin{equation}
\sigma_i^x = e^{-i\pi/2}R_i^x(\pi),
\end{equation}
where $2\delta E_J\tau/\hbar=\pi$.  Note that when the flux in the SQUID loop of the $i$th qubit is switched off to produce a local
$x$-rotation, the flux in the SQUID loop of the nearest-neighbor qubit that is connected to the $i$th qubit via an $LC$ oscillator should be
simultaneously shifted to a value around $\Phi_0/2$, so as to keep $|J_y|$ between these two qubits much smaller than $E_J$.

With both $\sigma_i^z$ and $\sigma_i^x$ rotations available for the $i$th qubit, the $\sigma_i^y$ rotation is given by
\begin{equation}
\sigma_i^y=e^{-i\pi/2}\sigma_i^z\sigma_i^x.
\end{equation}
Therefore, one can construct all the wanted operations $\tilde{\sigma}_i^z$ and $\tilde{\sigma}_i^z$ for generating both vortex and bond-state
excitations by using the single-qubit rotations $\sigma_i^z$ and $\sigma_i^x$.

In order to obtain accurate $z$- and $x$-type single-qubit rotations, we assume that $E_c$ and $E_J$ are much larger than the inter-qubit
coupling.  Actually this somewhat stringent condition can be loosened for realistic systems.  As shown in Ref.~\onlinecite{Wei}, accurate
effective single-qubit rotations can still be achieved using techniques from nuclear magnetic resonance when the inter-qubit coupling is small
compared to single-qubit parameters (instead of much smaller than $E_c$ and $E_J$).

\section{Discussion and conclusion}

In this paper, our main objective is to construct an experimentally
feasible proposal to emulate the Kitaev model on a network made of
superconducting nanocircuits.  To focus on the topological
properties of the system, we choose the limit of identical qubits
and identical coupling strength. Furthermore, we fix the mutual
inductances and the capacitances of the various circuit elements
involved.  There are basically two tunable parameters: the gate
voltage on the Cooper pair boxes ($n_g$) for each charge qubit, and
the magnetic flux $\Phi_e$ through the SQUID loops connected to the
Cooper pair boxes.  Within the regime where $z$-bonds dominate in
interaction energy scale ($J_z$ much larger than all other
couplings, including $J_x$, $J_y$, $h_x$, and $h_z$), we have
explored two limiting cases: one with weak effective magnetic fields
($|h_x|$, $|h_z|$ $\ll$ $J_x$, $|J_y|$), the other with the
effective field only along the $x$-direction.  We have identified
some properties of the relevant ground states, and the low-energy
excitations, the vortex and bond states.  However, much more study
is needed to completely clarify the energy spectrum, the phase
diagram, and the dynamics of this superconducting network.

One observation we have made is that the vortex excitations and bond-state excitations can be generated using the same spin-pair operations,
starting from different ground states ($|g\rangle_w$ and $|g\rangle_b$) that depend on the system parameters.  We have also shown that while
$|g\rangle_w$ is highly entangled, $|g\rangle_b$ is only entangled locally but not globally.  This quantum phase transition requires more
extensive studies to identify the critical point and related critical phenomena, such as how system entanglement changes near the transition
point, and most importantly how its topological properties change. It would also be worthwhile to investigate the system spectrum (from vortex
excitation to bond state excitation) and dynamics during this transition, similar to our study of quantum phase transitions between Abelian and
non-Abelian phases of the Kitaev model.\cite{Shi_PRB09} While such studies are generally numerically intensive, it would help reveal the exotic
topological properties of this many-body model.

With the elementary building blocks given in Fig.~\ref{fig1}(a), one can construct Kitaev spin models on {\it other} types of lattices as well
(see, e.g., Refs.~\onlinecite{Yao} and \onlinecite{Sun}).  In particular, it has been shown that in the absence of a magnetic field, the Kitaev
model on a decorated honeycomb lattice~\cite{Yao} can support gapped non-Abelian anyons. The quantum analog simulation of Kitaev models on
different lattices using superconducting circuits should shed light on the novel properties of these topological systems.
%

There are two important open issues in the study of building a
superconducting qubit network to emulate a spin lattice.  One is the
role played by the decoherence of individual qubits, and the other
is the measurement of correlated states on a qubit network.  It is
well known that charge qubits suffer from fast decoherence. However,
it is not clear how decoherence would affect the topological
excitations.  Indeed, the faster decoherence of charge qubits may
allow the system to reach its ground state faster. Furthermore,
topological excitations are supposed to be robust against local
fluctuations, so that decoherence in individual nodes may not easily
destroy excitations such as the vortex state.  Quantum measurement
is another open issue in the study of collective states, whether
ground states or low-energy excitations, of a qubit lattice. While
single-qubit measurement of superconducting qubits can now be done
with quite high fidelity,\cite{Holfheinz_Nature,DiCarlo_Nature} and
two-qubit correlation measurements have been done,\cite{Bell_UCSB}
measuring multi-qubit correlations requires further theoretical and
experimental studies.  We hope that our proposal acts as another
incentive for researchers in the field of superconducting qubits to
look for ways to perform measurements that can reveal quantum
correlations.

In conclusion, we have proposed an approach to emulate the Kitaev model on a honeycomb lattice using superconducting quantum circuits, and shown
that the low-energy dynamics of the superconducting network should follow a finite-field Kitaev model Hamiltonian.  We analytically study two
particular limits for system parameters, explore their ground state characteristics, and identify their low-energy excitations as vortex states
and bond states.  We further show that both vortex- and bond-state excitations can be generated using the same spin-pair operations, starting
from different ground states. Our proposal points to an experimentally realizable many-body system for the quantum emulation of the Kitaev
honeycomb spin model.

\begin{acknowledgments}
We thank J. Vidal and Yong-Shi Wu for useful discussions. J.Q.Y. and
X.F.S. were supported by the National Basic Research Program of
China Grant Nos. 2009CB929300 and 2006CB921205, and the National
Natural Science Foundation of China Grant Nos. 10625416 and
10534060.  X.H. and F.N. acknowledge support by the National
Security Agency and the Laboratory for Physical Sciences through the
US Army Research Office, X.H. acknowledges support and hospitality
by the Kavli Institute of Theoretical Physics at the University of
California at Santa Barbara, and F.N. thanks support by the National
Science Foundation Grant No.~0726909.
\end{acknowledgments}

\appendix
\section{Derivation of the Hamiltonian}

Below we derive the Hamiltonian of the honeycomb lattice constructed
with superconducting quantum circuits as described in
Fig.~\ref{fig1}.  For simplicity all charge qubits have the same
parameters.  Furthermore, the mutual inductances $M$, the $LC$
oscillators, and the mutual capacitances $C_m$ for the $x$-, $y$-,
and $z$-couplings are also identical, respectively.  Since the self
inductance of the SQUID loop in each charge qubit is small, the
voltage drop across this loop inductance can be ignored as compared
with the voltage drops across the Josephson junctions in the loop.
Also, we assume that the capacitance of the $LC$ oscillator is much
larger than the gate capacitance and the mutual capacitance, i.e.,
$C \gg C_g,C_m$.  The total electrical energy of the qubit lattice
can be written as (the 1-2 ad 1-3 couplings are magnetic and will be
discussed later)
\begin{equation}
T = \sum_{z-{\rm link}}T_{14},
\end{equation}
where the summation is over all the $z$-links. the term $T_{14}$
contains the charging energies of the nodes on either end of a
$z$-link in the building block, together with the coupling across
the link. It is given by
\begin{eqnarray}
T_{14}\!&\!=\!&\!\sum_{i=1,4} \frac{1}{2}C_{\Sigma}\left(\frac{\Phi_0}{2\pi}\right)^2\left[\dot{\varphi}_i^2
+ 2\left( \frac{2\pi}{\Phi_0}\right)\frac{\dot{a}_i+C_gV_{gi}}{C_{\Sigma}}\dot{\varphi}_i\right]\nonumber\\
&&\!-C_m\left(\frac{\Phi_0}{2\pi}\right)^2\dot{\varphi}_1\dot{\varphi}_4
+ \frac{1}{2}C\dot{\Phi}_L^2,
\end{eqnarray}
where
\begin{eqnarray}
\dot{a}_1\!&\!=\!&\!(C_g+C_m)\dot{\Phi}_L-C_m\dot{\Phi}_{L'},
\nonumber\\
\dot{a}_4\!&\!=\!&\!(C_g+C_m)\dot{\Phi}_{L'}-C_m\dot{\Phi}_L,
\end{eqnarray}
and $C_{\Sigma}=2C_J + C_g + C_m$.  Here $(\Phi_0/2\pi)\dot{\varphi}_i \equiv V_{Ji}$ is the average voltage drop across the two Josephson
junctions of the $i$th charge qubit, and $\dot{\Phi}_L\equiv V_L$ ($\dot{\Phi}_{L'}\equiv V_{L'}$) is the voltage drop across the $LC$
oscillator connected to qubit 1 (4).

The Langrangian of the qubit lattice is
\begin{equation}
\mathcal{L}=T-U \;,
\end{equation}
where $U$ is the total potential energy of the system, including Josephson coupling energy and magnetic energy in all the inductors in the
network.  To derive the system Hamiltonian, we choose $\varphi_i$ and $\Phi_L$ as the canonical coordinates.  The corresponding canonical
momenta are thus
\begin{eqnarray}
p_i\!&\!=\!&\!\frac{\partial L}{\partial\dot{\varphi}_i},\nonumber\\
p_L\!&\!=\!&\!\frac{\partial L}{\partial\dot{\Phi}_L}.
\end{eqnarray}
More explicitly,
\begin{eqnarray}
p_1\!&\!=\!&\!C_{\Sigma}\left(\frac{\Phi_0}{2\pi}\right)^2\dot{\varphi}_1
-C_m\left(\frac{\Phi_0}{2\pi}\right)^2\dot{\varphi}_4 \nonumber\\
&&+\left(\frac{\Phi_0}{2\pi}\right)(\dot{a}_1+C_gV_{g1}),\nonumber\\
p_4\!&\!=\!&\!C_{\Sigma}\left(\frac{\Phi_0}{2\pi}\right)^2\dot{\varphi}_4
-C_m\left(\frac{\Phi_0}{2\pi}\right)^2\dot{\varphi}_1 \nonumber\\
&&+\left(\frac{\Phi_0}{2\pi}\right)(\dot{a}_4+C_gV_{g4}),\nonumber\\
p_L\!&\!=\!&\!C\dot{\Phi}_L-C_m\left(\frac{\Phi_0}{2\pi}\right)(\dot{\varphi}_4+\dot{\varphi}_5)
\nonumber\\
&&+(C_g+C_m)\left(\frac{\Phi_0}{2\pi}\right)(\dot{\varphi}_1+\dot{\varphi}_2),
\end{eqnarray}
where the subscript 5 denotes the qubit which is connected to qubit
2 via the mutual capacitance $C_m$. In the limit of $C \gg C_g,C_m$,
$p_L \approx C\dot{\Phi}_L$. Thus one has
\begin{eqnarray}
\dot{\varphi}_1\!&\!=\!&\!\frac{C_{\Sigma}X_1+C_mX_4}{\left({\Phi_0}/{2\pi}\right)^2\Lambda},\nonumber\\
\dot{\varphi}_4\!&\!=\!&\!\frac{C_mX_1+C_{\Sigma}X_4}{\left({\Phi_0}/{2\pi}\right)^2\Lambda},\nonumber\\
\dot{\Phi}_L\!&\!=\!&\!\frac{p_L}{C},
\end{eqnarray}
with $\Lambda=C_{\Sigma}^2-C_m^2$, and
\begin{eqnarray}
X_1\!&\!=\!&\!p_1-\left(\frac{\Phi_0}{2\pi}\right)(\dot{a}_1+C_gV_{g1}),\nonumber\\
X_4\!&\!=\!&\!p_4-\left(\frac{\Phi_0}{2\pi}\right)(\dot{a}_4+C_gV_{g4}).
\end{eqnarray}

The Hamiltonian of the honeycomb lattice is thus
\begin{eqnarray}
H\!&\!=\!&\!\sum_{z-{\rm link}}(p_1\dot{\varphi}_1+p_4\dot{\varphi_4}+p_L\dot{\Phi}_L)-\mathcal{L} \nonumber\\
\!&\!=\!&\!\sum_{z-{\rm link}}\mathcal{T}_{14}+U,
\end{eqnarray}
where
\begin{equation}
\mathcal{T}_{14}=\frac{C_{\Sigma}X_1^2}{2\left(\frac{\Phi_0}{2\pi}\right)^2\Lambda} + \frac{C_{\Sigma}X_4^2}{2 \left( \frac{\Phi_0}{2\pi}
\right)^2\Lambda} + \frac{C_mX_1X_4}{\left( \frac{\Phi_0}{2\pi} \right)^2\Lambda} + \frac{p_L^2}{2C}.
\end{equation}
We now perform two gauge transformations, so that $p_1$ and $p_4$ become:
\begin{eqnarray}
p_1-\left(\frac{\Phi_0}{2\pi}\right)\dot{a}_1=\widetilde{p}_1,\nonumber\\
p_4-\left(\frac{\Phi_0}{2\pi}\right)\dot{a}_4=\widetilde{p}_4 \,.
\end{eqnarray}
After these gauge transformations, $\dot{\varphi}_1$ and
$\dot{\varphi}_4$ become
\begin{eqnarray}
\dot{\varphi}_1-\frac{C_{\Sigma}}{\Lambda}\left(\frac{2\pi}{\Phi_0}\right)\dot{a}_1
= \dot{\widetilde{\varphi}}_1,\nonumber\\
\dot{\varphi}_4-\frac{C_{\Sigma}}{\Lambda}\left(\frac{2\pi}{\Phi_0}\right)\dot{a}_4
=\dot{\widetilde{\varphi}}_4,
\end{eqnarray}
and $\mathcal{T}_{14}$ can be expressed as
\begin{eqnarray}
\mathcal{T}_{14}\!&\!=\!&\!\frac{1}{2}C_{\Sigma}\mathcal{K}^2(\widetilde{p}_1,V_{g1})
+\frac{1}{2}C_{\Sigma}\mathcal{K}^2(\widetilde{p}_4,V_{g4})\nonumber\\
&&\!+C_m\mathcal{K}(\widetilde{p}_1,V_{g1})\mathcal{K}(\widetilde{p}_4,V_{g4})+\frac{p_L^2}{2C},
\end{eqnarray}
where
\begin{equation}
\mathcal{K}(\widetilde{p}_i,V_{gi})=\frac{\widetilde{p}_i-(\frac{\Phi_0}{2\pi})C_gV_{gi}}
{\left(\frac{\Phi_0}{2\pi}\right)\Lambda^{1/2}}. \label{mathcalk}
\end{equation}
Based on these building blocks, instead of the $z$-links, the
Hamiltonian of the qubit lattice can now be rewritten as
\begin{equation}
H=\sum_{\rm BB} \mathcal{T}_{\rm BB} + U.
\end{equation}
Here the summation is over all the building blocks and
$\mathcal{T}_{\rm BB}$, for a building block shown in
Fig.~\ref{fig1}(a), is given by
\begin{eqnarray}
\mathcal{T}_{\rm
BB}\!&\!=\!&\!\frac{1}{2}C_{\Sigma}\mathcal{K}^2(\widetilde{p}_1,V_{g1})
+\frac{1}{6}C_{\Sigma}\mathcal{K}^2(\widetilde{p}_2,V_{g2})\nonumber\\
&&\!+\frac{1}{6}C_{\Sigma}\mathcal{K}^2(\widetilde{p}_3,V_{g3})
+\frac{1}{6}C_{\Sigma}\mathcal{K}^2(\widetilde{p}_4,V_{g4})\nonumber\\
&&\!+C_m\mathcal{K}(\widetilde{p}_1,V_{g1})\mathcal{K}(\widetilde{p}_4,V_{g4})
+\frac{p_L^2}{2C}, \label{kinetic}
\end{eqnarray}
where the prefactor $\frac{1}{6}$ in the second, third and fourth terms (instead of $\frac{1}{2}$ as in the first term) is due to the lattice
geometry that each of the qubits 2-4 is shared by three building blocks.

The potential energy of the system consists of the Josephson energy
$-E_{Ji}(\Phi_{iL})\cos\varphi_i$ of each qubit, the magnetic energy
$\Phi_L^2/2L$ of each $LC$ oscillator, the self-inductance energy
$\frac{1}{2} L_q I_i^2$ of each qubit, and the mutual-inductance
energy $-MI_iI_j$ between every pair of nearest-neighbor qubits
coupled via $M$.  In particular, the Josephson coupling energy is
\begin{equation}
E_{Ji}(\Phi_{iL})=2E_J\cos\left(\frac{\pi\Phi_{iL}}{\Phi_0}\right) \,.
\label{energy}
\end{equation}
The supercurrent in the SQUID loop of the $i$th qubit is
\begin{equation}
I_i=-I_c\sin\left(\frac{\pi\Phi_{iL}}{\Phi_0}\right)\cos\varphi_i \,.
\label{current}
\end{equation}
Here $I_c=2\pi E_J/\Phi_0$ is the critical current of the Josephson junction, $\Phi_0=h/2e$ is the flux quantum, $L_q$ is the SQUID loop
inductance of each qubit, and the total magnetic flux $\Phi_{iL}$ in the loop of qubit $i$ is given by
\begin{equation}
\Phi_{iL}=\Phi_i + L_q I_i - MI_j,
\end{equation}
with $\Phi_i$ the externally applied magnetic flux in the loop of qubit $i$ and $I_j$ is the supercurrent in the loop of qubit $j$ that is
coupled to qubit $i$ via $M$. Based on the building blocks, the potential energy $U$ can been written as
\begin{equation}
U=\sum_{\rm BB}U_{\rm BB},
\end{equation}
with
\begin{eqnarray}
U_{\rm BB}\!&\!=\!&\!-E_{J1}(\Phi_{1L})\cos\varphi_1
-\frac{1}{3}E_{J2}(\Phi_{2L})\cos\varphi_2 \nonumber\\
&&\!-\frac{1}{3}E_{J3}(\Phi_{3L})\cos\varphi_3
-\frac{1}{3}E_{J4}(\Phi_{4L})\cos\varphi_4 \nonumber\\
&&\!+\frac{\Phi_L^2}{2L}+\frac{1}{2}L_qI_1^2+\frac{1}{6}L_qI_2^2
+\frac{1}{6}L_qI_3^2\nonumber\\
&&\!+\frac{1}{6}L_qI_4^2-MI_1I_2,
\end{eqnarray}
where the prefactors $\frac{1}{3}$ and $\frac{1}{6}$ are again due
to the lattice geometry that each of the qubits 2-4 is shared by
three building blocks.

Usually, the self inductance $L_q$ and the mutual inductance are much smaller than the Josephson inductance of each junction in the qubit loop.
Thus, one can expand Eqs.~(\ref{energy}) and (\ref{current}) around $\pi\Phi_i/\Phi_0$ and keep the leading terms, as in Ref.~\onlinecite{YTN}.
The potential energy is then reduced to
\begin{eqnarray}
U_{\rm BB}\!&\!=\!&\!-E_{J1}(\Phi_{1})\cos\varphi_1 -\frac{1}{3}E_{J2}(\Phi_{2})\cos\varphi_2
\nonumber\\
&&\!-\frac{1}{3}E_{J3}(\Phi_{3})\cos\varphi_3 -\frac{1}{3}E_{J4}(\Phi_{4})\cos\varphi_4 \nonumber\\
&&\!+\frac{\Phi_L^2}{2L}+MI_1I_2,
\label{potential}
\end{eqnarray}
where the supercurrents $I_i$ are replaced by
\begin{equation}
I_i=-I_c\sin\left(\frac{\pi\Phi_{i}}{\Phi_0}\right)\cos\varphi_i.
\end{equation}
In Eq.~(\ref{potential}), we have also omitted constant terms which are reduced to identity operators in the qubit subspace, because these terms
only shift the zero energy of the system.

Using the gauge transformation
\begin{equation}
\varphi_i = \widetilde{\varphi}_i+ \frac{C_{\Sigma}}{\Lambda}\left(\frac{2\pi}{\Phi_0}\right)a_i,
\end{equation}
when the fluctuations of $a_i$ are weak so that~\cite{Mak}
\begin{equation}
\frac{C_{\Sigma}}{\Lambda}\sqrt{\langle a^2\rangle}\ll\Phi_0,
\end{equation}
one has
\begin{equation}
\cos\varphi_i\approx\cos\widetilde{\varphi}_i
-\left[\frac{C_{\Sigma}}{\Lambda}
\left(\frac{2\pi}{\Phi_0}\right)a_i\right]\sin\widetilde{\varphi}_i.
\end{equation}
The potential energy $U_{BB}$ is given by
\begin{eqnarray}
U_{\rm BB}\!&\!=\!&\!-E_{J1}(\Phi_{1})\cos\widetilde{\varphi}_1
-\frac{1}{3}E_{J2}(\Phi_{2})\cos\widetilde{\varphi}_2\nonumber\\
&&\!-\frac{1}{3}E_{J3}(\Phi_{3})\cos\widetilde{\varphi}_3
-\frac{1}{3}E_{J4}(\Phi_{4})\cos\widetilde{\varphi}_4\nonumber\\
&&\!+\frac{1}{2L}\left[\Phi_L+\left(\frac{2\pi L}{\Phi_0}\right)
(Y_1+Y_2)\right]^2\nonumber\\
&&\!-\left(\frac{2\pi^2L}{\Phi_0^2}\right)(Y_1+Y_2)^2+MI_1I_2,
\label{UBB}
\end{eqnarray}
with
\begin{equation}
Y_i=\frac{C_{\Sigma}(C_g+C_m)}{\Lambda}E_{Ji}(\Phi_i)\sin\widetilde{\varphi}_i.
\end{equation}
Here the terms modifying the Josephson coupling energy are ignored because they are much smaller than the Josephson coupling energy.

The term ${p_L^2}/{2C}$ in Eq.~(\ref{kinetic}) is the kinetic energy of the $LC$ oscillator and the term
$\frac{1}{2L}\left[\Phi_L+\left(\frac{2\pi L}{\Phi_0}\right) (Y_1+Y_2)\right]^2$ in Eq.~(\ref{UBB}) is the potential energy of the $LC$
oscillator. When the frequency of the $LC$ oscillator is much larger than the qubit frequency, the $LC$ oscillator remains in the ground state,
so that these terms can be removed from the Hamiltonian.~\cite{Mak} Thus, the Hamiltonian of the system can finally be written as
\begin{eqnarray}
H\!&\!=\!&\!\sum_i H_i + \sum_{x-{\rm link}}K_x(j,k) + \sum_{y-{\rm
link}}K_y(j,k) \nonumber \\
&&\!+\sum_{z-{\rm link}}K_z(j,k). \label{Ham}
\end{eqnarray}
Here
\begin{equation}
H_i=\frac{C_{\Sigma}}{2}\mathcal{K}^2(\widetilde{p}_i,V_{gi})
-E_{Ji}(\Phi_i)\cos\widetilde{\varphi}_i,
\end{equation}
with $\mathcal{K}(\widetilde{p}_i,V_{gi})$ given in Eq.~(\ref{mathcalk}).  For the building block shown in Fig.~{\ref{fig1}(a), the three
nearest-neighbor couplings $K_x$, $K_y$ and $K_z$ are given by
\begin{eqnarray}
K_x(1,2)\!&\!=\!&\!MI_1I_2, \nonumber\\
K_y(1,3)\!&\!=\!&\!-4\xi
E_{J1}(\Phi_1)E_{J3}(\Phi_3)\sin\widetilde{\varphi}_1\sin\widetilde{\varphi}_3,\nonumber\\
K_y(1,4)\!&\!=\!&\!C_m\mathcal{K}(\widetilde{p}_1,V_{g1})\mathcal{K}(\widetilde{p}_4,V_{g4}),
\end{eqnarray}
where
\begin{equation}
\xi=L \left[\frac{\pi C_{\Sigma} (C_g +
C_m)}{\Lambda\Phi_0}\right]^2.
\end{equation}
In Eq.~(\ref{Ham}), the terms with $\sin^2\widetilde{\varphi}_i$ are also removed because they are reduced to the identity operators in the
qubit subspace. The canonical coordinates $\widetilde{\varphi}_i$ and momenta $\widetilde{p}_i$ are conjugate variables, and they obey the
commutation relation:
\begin{equation}
[\widetilde{\varphi}_j,\widetilde{p}_k]=i\hbar\delta_{jk},
\end{equation}
where
$\widetilde{p}_j=-i\hbar\partial/\partial\widetilde{\varphi}_j$.
Defining $\widetilde{n}_i\equiv\widetilde{p}_i/\hbar$, one obtains
Eq.~(\ref{hamiltonian}) by replacing $\widetilde{n}_i$ and
$\widetilde{\varphi}_i$ in Eq.~(\ref{Ham}) with $n_i$ and
$\varphi_i$.

Below we give two examples of parameter regimes where the physics we
discussed in this paper can be realized.  For a quantum circuit with
two charge qubits coupled by a mutual capacitance, the typical
parameters are $C_J \approx 500$~aF, $C_m \approx 30$~aF, $C_g
\approx 0.5$~aF, and $E_J \approx 15$~GHz (see, e.g.,
Ref.~\onlinecite{NEC}).  Here we choose $C_m = 200$~aF so as to have
a stronger capacitive coupling, $C_J \approx 400$~aF, and $C_g
\approx 0.5$~aF.  These parameters give $E_c\approx 80$~GHz and
$J_z\approx 8$~GHz.  We also choose $E_J = 20$~GHz and apply a
magnetic flux $\Phi_e$ in each qubit loop such that
$E_J(\Phi_e)\approx 4$~GHz.  This gives $|h_x|\approx 2$~GHz.
Because $h_z$ can be independently controlled by the gate voltage,
it is easy to obtain $|h_z|\approx |h_x|$.  Finally, we choose
$M\approx 6.6$~nH and the parameters of the $LC$ oscillator are
chosen as $L\approx 3.8$~$\mu$H and $C=4C_m=800$~aF.  We then have
$J_x\approx |J_y|\approx 4$~GHz.  The parameter regime given in
Sec.~III.A (i.e., $J_z \gg J_x, |J_y| \gg |h_z|, |h_x|$) can thus be
approximately achieved.  Also, $J_z$ is much smaller than the
frequency of the $LC$ oscillator $\omega=1/\sqrt{LC}\approx 20$~GHz,
so that the lattice dynamics can be reasonably described by the
Kitaev model in this regime.  Note that $E_c\approx 80$~GHz and
$E_J=20$~GHz, which are much larger than $J_z$.  Thus, the local
quantum rotations $\sigma_i^z$ and $\sigma_i^x$ for generating
topological excitations can also be achieved.  Though $E_c$ and
$E_J$ are much larger than or comparable to the frequency of the
$LC$ oscillator, the local quantum rotations are implemented by
changing the external fields applied locally on the qubits involved.
It is expected that the total Kitaev lattice will not be affected so
much by these local operations because the topological properties
should be robust against local fluctuations.

For the parameter regime of Sec.~III.B, we choose $L\approx
2.5$~$\mu$H, $M\approx 4.9$~nH, and $n_g=\frac{1}{2}$.  The applied
magnetic flux in each qubit loop is such that $E_J(\Phi_e)=3$~GHz.
Other system parameters are chosen to be the same as those in the
case above.  Thus, we have $J_z \approx 8$~GHz, $J_x \approx |J_y|
\approx |h_x| \approx 3$~GHz, and $|h_z|=0$.  These parameters are
much smaller than the frequency of the $LC$ oscillator $\omega
\approx 20$~GHz, allowing us to consider only the ground state of
the oscillator.  Thus the Kitaev lattice can also be realized in
this regime.  Moreover, because $E_c \approx 80$~GHz and $E_J =
20$~GHz, which are much larger than $J_z$, the local quantum
rotations $\sigma_i^z$ and $\sigma_i^x$ at the $i$th site can be
implemented.

With the parameters considered here, the vortex excitation energy
would be of the order of 0.1 GHz or larger, corresponding to an
experimental temperature of 10 mK or higher, already accessible by
currently available dilution refrigerators.


\end{document}